\documentclass[sn-mathphys, Numbered, iicol]{sn-jnl}

\usepackage{graphicx}%
\usepackage{multirow}%
\usepackage{amsmath,amssymb,amsfonts}%
\usepackage{amsthm}%
\usepackage{mathrsfs}%
\usepackage[title]{appendix}%
\usepackage{xcolor}%
\usepackage{textcomp}%
\usepackage{manyfoot}%
\usepackage{booktabs}%
\usepackage{algorithm}%
\usepackage{algorithmicx}%
\usepackage{algpseudocode}%
\usepackage{listings}%
\usepackage{cleveref}
\usepackage{ulem}
\usepackage{xspace}
\usepackage{mhchem}
\usepackage{subcaption}
\usepackage{bbm}
\usepackage{placeins}
\makeatletter

\makeatletter
\newcommand*{\myfnsymbolsingle}[1]{%
  \ensuremath{%
    \ifcase#1%
    \or %
      \S%
    \or %
      \dagger
    \or %
      a
    \or %
      b
    \or %
      c
    \or 
      d
    \else %
      \@ctrerr  
    \fi
  }%
}   
\makeatother
\usepackage{alphalph}
\usepackage{fnpct} %
\usepackage{xr}

\theoremstyle{thmstyleone}%

\theoremstyle{thmstyletwo}%

\theoremstyle{thmstylethree}%

\Crefname{figure}{Fig.}{Figs.}
\Crefname{section}{Section}{Sections}
\Crefname{equation}{Eq.}{Eqs.}

\Crefname{algorithm}{Algorithm.}{Algorithms.}

\newcommand{\nameframework}{SAAM\xspace}
\newcommand{\namenetwork}{SA-LapNet\xspace}

\newcommand{\spnote}[1]{Supplementary Note #1}

\usepackage{amsmath,amsfonts,bm}

\def\eqref#1{equation~\ref{#1}}

\def\1{\bm{1}}

\def\vf{{\bm{f}}}
\def\vg{{\bm{g}}}

\def\vo{{\bm{o}}}

\def\vr{{\bm{r}}}

\def\vx{{\bm{x}}}

\def\mR{{\bm{R}}}

\DeclareMathAlphabet{\mathsfit}{\encodingdefault}{\sfdefault}{m}{sl}
\SetMathAlphabet{\mathsfit}{bold}{\encodingdefault}{\sfdefault}{bx}{n}

\newcommand{\ket}[1]{|{#1}\rangle}
\newcommand{\bra}[1]{\langle{#1}|}
\newcommand{\braket}[2]{\langle{#1}|{#2}\rangle}

\newcommand{\mbi}[0]{\mathbbm{i}}
\newcommand{\Npara}[0]{N_{\text{par}}}

\begin{document}

\title[Article Title]{Spin-Adapted Neural Network Wavefunctions in Real Space}

\author*[1,2]{Ruichen Li}
\equalcont{These authors contributed equally to this work.}
\email{lrc@bytedance.com}

\author[1]{Yuzhi Liu}\nomail
\equalcont{These authors contributed equally to this work.}

\author[1,2]{Du Jiang}\nomail
\author[1]{Yixiao Chen}\nomail
\author[1]{Xuelan Wen}\nomail
\author[1,2]{Wenrui Li}\nomail

\author*[2]{Di He}
\email{dihe@pku.edu.cn}

\author*[2]{Liwei Wang}
\email{wanglw@pku.edu.cn}

\author*[2]{Ji Chen}
\email{ji.chen@pku.edu.cn}

\author*[1]{Weiluo Ren}
\email{renweiluo@bytedance.com}

\affil[1]{\orgname{ByteDance Seed}}
\affil[2]{\orgname{Peking University}}

\abstract{
Spin plays a fundamental role in understanding electronic structure, yet many real-space wavefunction methods fail to adequately consider it.
We introduce the Spin-Adapted Antisymmetrization Method (SAAM), a general procedure that enforces exact total spin symmetry for antisymmetric many-electron wavefunctions in real space.
In the context of neural network-based quantum Monte Carlo (NNQMC), SAAM leverages the expressiveness of deep neural networks to capture electron correlation while enforcing exact spin adaptation via group representation theory.
This framework provides a principled route to embed physical priors into otherwise black-box neural network wavefunctions, yielding a compact representation of correlated system with neural network orbitals.
Compared with existing treatments of spin in NNQMC, \nameframework is more accurate and efficient, achieving exact spin purity without any additional tunable hyperparameters.
To demonstrate its effectiveness, we apply \nameframework to study the spin ladder of iron-sulfur clusters, a long-standing challenge for many-body methods due to their dense spectrum of nearly degenerate spin states. 
Our results reveal accurate resolution of low-lying spin states and spin gaps in \ce{[Fe2S2]} and \ce{[Fe4S4]} clusters, offering new insights into their electronic structures.
In sum, these findings establish \nameframework as a robust, hyperparameter-free standard for spin-adapted NNQMC, particularly for strongly correlated systems.
}

\maketitle

\small

\section{Main}\label{sec: Main}

Accurately describing electron correlation remains a central challenge in quantum chemistry.
One major difficulty is the proper characterization of complex spin structure for strongly correlated systems such as correlated materials and transition metal complexes. 
Inadequate spin treatment prevents methods from correctly capturing static correlation, leading to critical issues like spurious low energy, misordered excitations, and misestimated barriers \cite{symmetry-dilemma-1,symmetry-dilemma-2}.
Consequently, spin must be put at the center of the formulation when targeting quantitative prediction and insight for those systems.

However, even state-of-the-art computational methods can get spin wrong.
For instance, coupled-cluster singles and doubles with perturbative triples (CCSD(T)), the golden standard method in quantum chemistry, can generate spin-contaminated states for open-shell systems.
Another notable example is neural network-based quantum Monte Carlo (NNQMC), where highly expressive  neural network wavefunctions \cite{carleo2017solving,ferminet,hermann2020deep} typically do not enforce exact spin symmetry. 
Existing remedies for NNQMC requires careful tuning of additional hyperparameters to balance optimization quality against residual spin contamination \cite{spluspenalty,szabo2024improved}, and this balance is often fragile.

In this work, we introduce a general procedure to encode spin information directly into the real-space wavefunction ans{\"a}tz.
Our approach, the Spin Adapted Antisymmetrization Method (SAAM), provides a practical spin-adaptation protocal based on a fundamental insight of the nonrelativistic electronic Hamiltonian, namely that its eigenstates factorize into independent spin and spatial components.
Specifically, SAAM enforces spin symmetry by constructing spin functions from group representation theory with no added hyperparameters, while preserving full expressiveness for spatial correlations.
By modeling the spatial component with powerful neural network orbitals (NNO), NNQMC provides the natural platform for SAAM to facilitate accurate spin-adapted solutions to the many-electron Schrödinger equation.
\nameframework further offers a chemical interpretation of NNO, naturally enabling the definition of chemical concepts such as core/active  within real-space NNQMC framework. 
This integration thus provides a compact, chemically interpretable representation of realistic electron wavefunctions, marking the step toward a deeper understanding of correlated systems.

To demonstrate effectiveness of our approach, we first calculate singlet-triplet gap for biradical systems and obtain highly accurate predictions.
Then we apply \nameframework to excite-state calculations by integrating with the natural excited states (NES) method \cite{NES}.
This combination is both efficient and accurate, as showcased on carbon dimer.
Leveraging these advances, we accurately characterize iron–sulfur clusters, notorious for their complex near-degenerate spin spectrum.
Our results underscore the transformative potential of explicitly embedding fundamental physical symmetries within neural network-based wavefunctions, opening new avenues to extend \textit{ab initio} simulations into strongly correlated and multireferencial regimes with unprecedented accuracy and reliability.

\section{Results}\label{sec: Results}
\subsection{Overview of \nameframework}\label{subsec: frameworks}
\begin{figure*}[ht]
    \centering
    \includegraphics[width=1.0\linewidth]{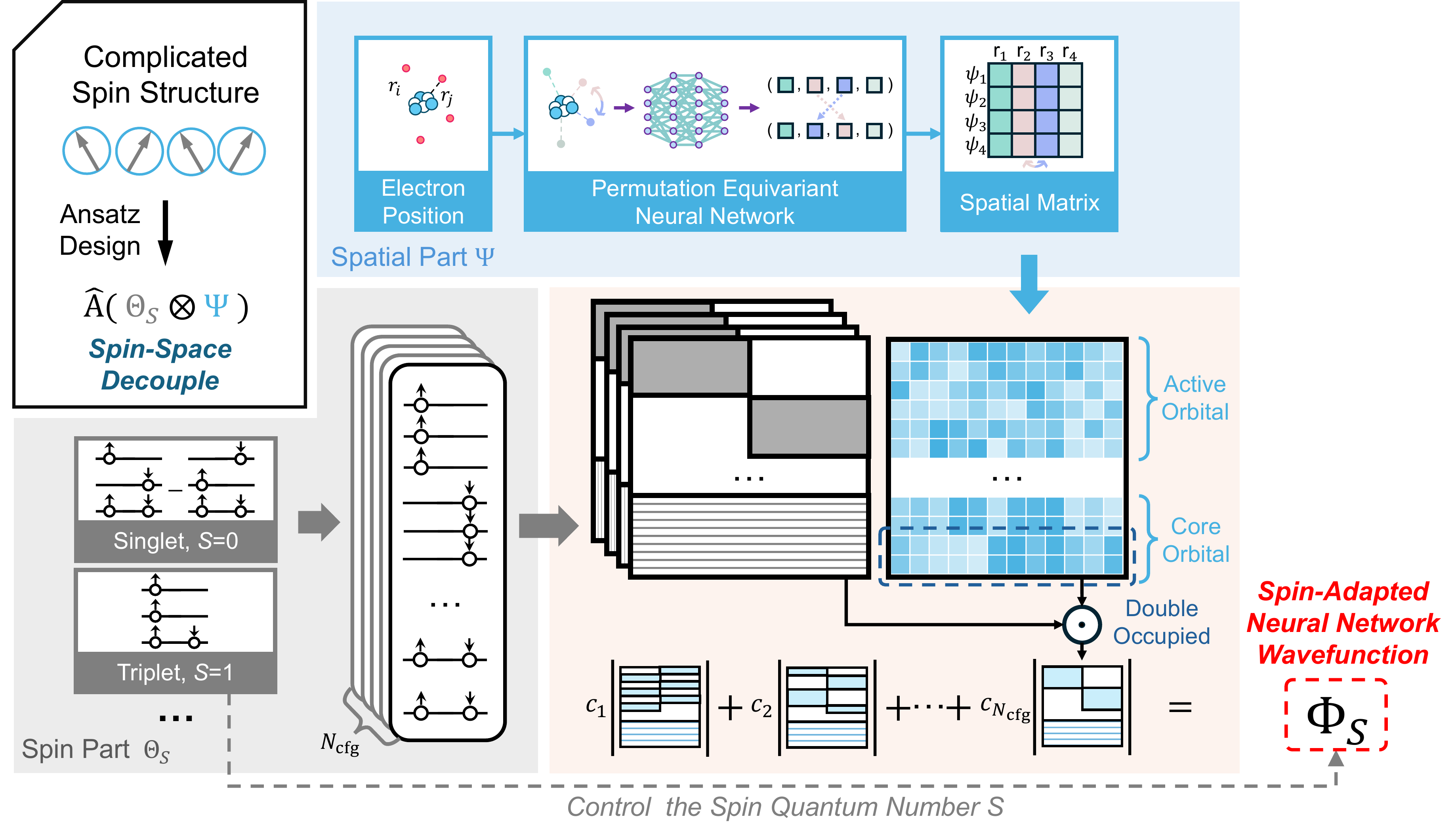}
     \caption{{\bf Overview of \nameframework.} We use a spin-spatial decoupled ans{\"a}tz to represent the complicated spin structure in molecular systems. Spatial part $\Psi$ (blue panel): Following the convention of continuous-space NNQMC, electron coordinates are processed by a permutation-equivariant neural network to capture many-body correlations. The network outputs the spatial matrix  $M_{\text{space}}$, where rows correspond to $\psi_i$ and columns correspond to electron positions. Spin part $\Theta_S$ (grey panel): The spin structure is assigned based on chemical prior knowledge and decomposed into products of one-body spin functions. For clarity, the decomposition is illustrated using standard spin-up/down functions.
      The spatial and spin parts are combined and antisymmetrized to yield a spin-adapted neural network wavefunction (bottom right, pink panel). Concretely, the spatial output is arranged into a matrix, in which the lowest $N_c$ rows (core orbitals) are duplicated to represent double occupied orbitals. Matrices derived from the spin decomposition and electron spin coordinates are then multiplied with the matrix from spatial part. Here, we use the dashed matrix in the spin part to represent the rows related to the core orbitals. The final spin-adapted neural network wavefunction is obtained as a weighted sum of determinants of these matrices, with coefficients determined by the spin decomposition. }
     \label{fig:framework}
\end{figure*}

We introduce the Spin-Adapted Antisymmetrization Method
(SAAM), which develops spin-adapted ans{\"a}tz for real-space quantum chemistry methods targeting the non-relativistic electronic Hamiltonian $\hat{H}$.
SAAM utilizes spin-spatial decoupled ans{\"a}tz 
\begin{align}
    \Phi=\hat{A}(\Theta_{S}\otimes\Psi),
    \label{eq: spin-spatial decouple}
\end{align}
where $\hat{A}$ is the antisymmetrization operator; $\Psi$ is a function in the spatial Hilbert space; $\Theta_S$ is a function in the spin Hilbert space with total spin quantum number $S$.
This separation of wavefunction is consistent with the eigenstate structure of the spin-independent Hamiltonian $\hat{H}$.
However, the antisymmetrization operator $\hat{A}$ involves a summation over all electron permutations, resulting in factorial computational complexity for the general forms of $\Theta_{S}$ and $\Psi$. 
Consequently, spin-adapted ans{\"a}tze have traditionally been limited to simple one-body functions \cite{paldus1979configuration} or to systems with few electrons \cite{huang1998spin}.

The core contribution in this study is to integrate the ans{\"a}tz form \cref{eq: spin-spatial decouple} with powerful neural networks, enabling accurate descriptions of electronic structures in systems exhibiting both strong static and dynamic correlation. The overall procedure is illustrated in \cref{fig:framework}. We model the spatial part $\Psi$ as:
\begin{equation}
\Psi(\vr_1,\vr_2,...,\vr_N)=\prod_{i=1}^{N}\psi_i(\vr_i|\{\vr\}),\label{eq: spatial multiply}
\end{equation}
where $N$ is the number of electrons. $\psi_i$ is a permutation equivariant function. $\{\vr\}=\{\vr_1,\vr_2,...,\vr_N\}$ denotes the set of all electron positions. In \nameframework, a permutation equivariant function $\psi_i$ may appear one or two times in the spatial part, corresponding to active or core orbitals, respectively. The relation between the occurrence of $\psi_i$ and the conventional concept of occupation is discussed in \cref{sec:NNO}.
In \nameframework, we decompose $\Theta_S$ into products of one-body spin functions:
\begin{equation}
\Theta_{S}=\sum_{t=1}^{N_{\text{cfg}}}c_t\chi_1^t(\sigma_1)\chi_2^t(\sigma_2)...\chi_N^{t}(\sigma_N), \label{eq: spin decomposition}
\end{equation}
where $N_{\text{cfg}}$ denotes the number of products required to represent the $\Theta_{S}$. $\chi_{i}^t:\{\frac{1}{2},-\frac{1}{2}\}\rightarrow\mathbb{C}$ is a normalized one-body spin function. 
$c_t\in \mathbb{C}$ is the coefficient corresponding to each product. $\sigma_i$ is the spin coordinate of the $i$-th electron. 

Combining spin and spatial parts, where  
$\psi_i(\vr_j|\{\vr\})$ and $\chi_i^t(\sigma_j)$ transform compatibly under electron permutations, we can compute the fully antisymmetric wavefunction as a weighted sum of determinants:
\begin{equation}\Phi(\vx_1,\vx_2,...,\vx_N)=\sum_{t=1}^{N_{\text{cfg}}}c_t\det[M^{t}], \label{eq: sum dets}
\end{equation}
where $\vx_j=(\vr_j,\sigma_j)$ represent the full coordinates of the $j$-th electron;  $M^{t}\in\mathbb{C}^{N\times N}$ is the Hadamard product of the spatial matrix $M_{\text{space}}$ and the spin matrices $M_{\text{spin}}^t$, where $[M_{\text{space}}]_{ij}=\psi_i(\vr_j|\{\vr\})$ and $[M_{\text{spin}}^t]_{ij}=\chi_i^t(\sigma_j)$. Following the convention in NNQMC, we assume that the first $N^{\uparrow}$ electrons are spin-up electrons and the last $N^{\downarrow}=N-N^{\uparrow}$ electrons are spin-down electrons, i.e., $\sigma_j=\frac{1}{2}$ for $j\leq N^{\uparrow}$ and $\sigma_j=-\frac{1}{2}$ for $N^{\uparrow}<j\leq N$. As the total spin angular momentum operator $\hat{S}^2$ commutes with the antisymmetrization operator, the resulting wavefunction $\Phi$ has the same total spin quantum number as $\Theta_S$, thereby providing a spin-adapted neural network wavefunction in real space.

\begin{figure*}[ht]
    \centering
    \includegraphics[width=\linewidth]{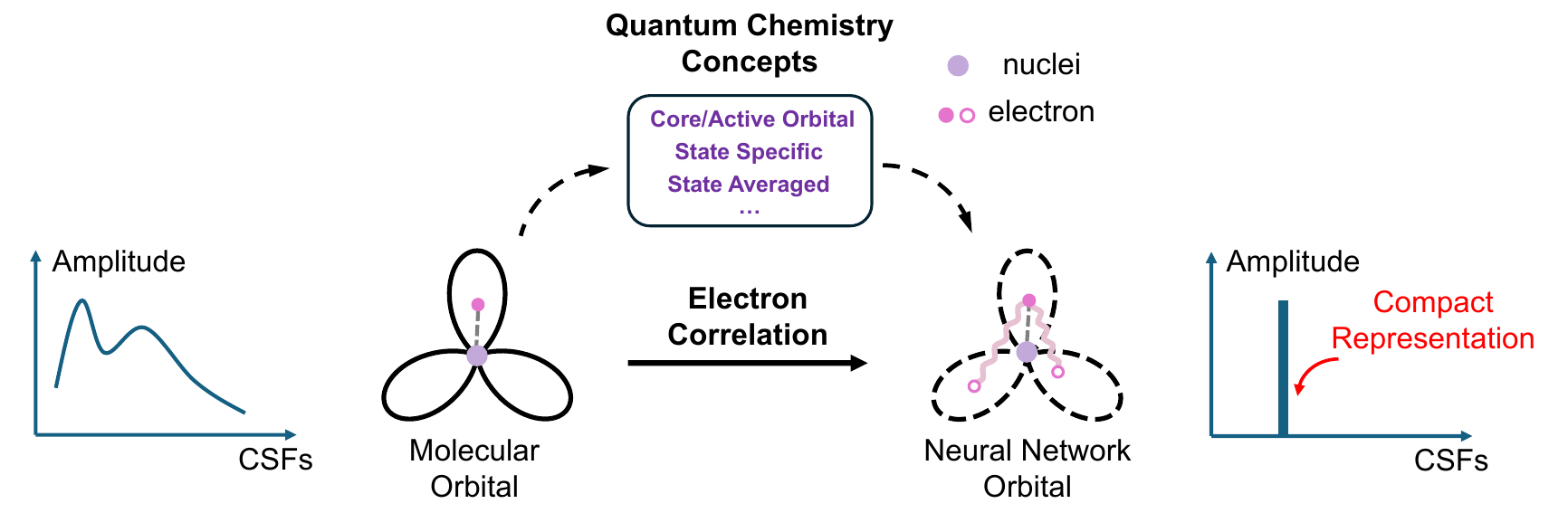}
     \caption{{\bf Classical quantum chemistry concepts extended to NNOs}. 
     NNOs allow natural extension of classical quantum chemistry concepts, including core/active orbitals, state-averaged, and state-specific excited-state calculations.     
     This bridges between chemical prior knowledge with NNQMC algorithms.
     Compared to one-body molecular orbitals, NNOs are able to capture electron correlations, enabling a compact, chemical-inspired representation of real-world electron wavefunction with fewer configuration state functions (CSF).
}
     \label{fig:concept}
\end{figure*}

\subsection{Neural Network Orbital}
\label{sec:NNO}
Beyond spin adaptation, \nameframework introduces a unified approach for integrating neural network wavefunctions with quantum chemical concepts by generalizing the notion of orbital occupation. Orbital occupation provides the heuristic foundation for interpreting electronic structure, where each orbital—defined as a one-body wavefunction—hosts up to two electrons in a non-interacting system. Electron correlation, however, often breaks this simple picture, motivating a broader definition of orbitals in advanced quantum chemistry.

Within SAAM, we establish a formal concept of Neural Network Orbitals (NNOs), a new class of generalized orbitals defined as permutation equivariant functions $\psi_i(\cdot|\{\mathbf r\})$.
Unlike its counterparts in existing NNQMC literature, the occupations of these orbitals, interpreted as occurrences in the spatial component of the wavefunction, carry explicit chemical meaning and align with the behavior of classical orbitals.
For example, the anti-symmetry restricts each NNO to one or two appearances, corresponding to single or double occupation. Moreover, for the double occupation, the associated spin component should form a singlet state (see \spnote{2}). Thus, the NNO provides a natural generalization of the classical orbitals to the correlated systems.

The extended orbital concept enables direct transfer of well-established ideas from traditional quantum chemistry to NNQMC methodologies, improving computational efficiency while maintaining high accuracy for strongly correlated systems. For instance, the familiar notions of core and active orbitals naturally extend to \nameframework, reducing the computational cost associated with inner-shell electrons. In \cref{fig:concept}, we further illustrate that several other quantum chemistry concepts can be seamlessly reformulated within the NNO framework.
For instance, we can generalize the idea of state-averaged and state-specific approaches in excited-state calculations. The state-averaged method uses a shared set of NNOs across multiple states to improve convergence, while the state-specific approach finetunes the orbitals for each individual state to provide accurate results.

Within our framework, neural network wavefunctions and established chemical understanding are mutually beneficial. 
On the one hand, the chemical prior knowledge inspires better wavefunction ansatz design and accelerates the NNQMC algorithms.
Conversely, NNOs yield a compact wavefunction representation for correlated systems, providing a deeper understanding of chemical systems with strong correlation effects and avoiding an extensive expansion of configuration state functions.

In the following sections, we demonstrate the effectiveness of \nameframework on chemical systems such as iron-sulfur clusters. 
Note that \nameframework is a general spin-adaptation protocol and can be applied to various real-space neural network ans{\"a}tz  \cite{ferminet,glehn2023a,li2024computational}.
Specifically, we use LapNet \cite{li2024computational} as the backbone network within \nameframework framework for all the calculation in this study (See \cref{alg:SA-LapNet}) and term the resulting ans{\"a}tz \namenetwork. 

\subsection{Performance Benchmark}\label{subsec:biradical}

\begin{figure*}[t]
    \centering
    \includegraphics[width=1.0\linewidth]{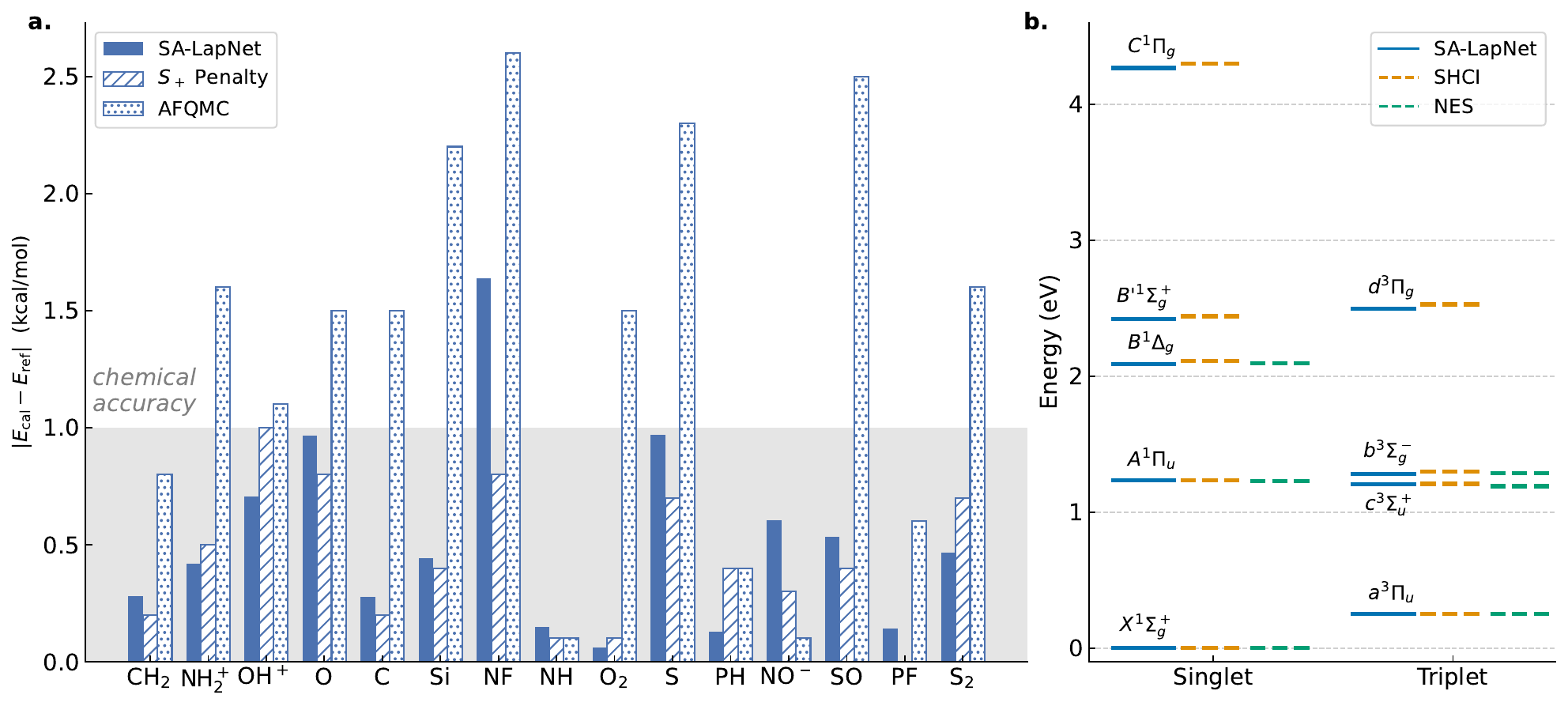}
     \caption{{\bf Benchmark of \namenetwork.} {\bf a}. Absolute deviations of calculated singlet-triplet (S-T) gaps are shown for a diverse set of biradical systems. 
     The reference experimental results are from Ref.\cite{biradical_shee,biradical_lee}.
     Our results (solid bars) are compared with $S_{+}$ penalty \cite{spluspenalty} (hatched bars), and auxiliary-field quantum Monte Carlo \cite{biradical_lee, biradical_shee} (AFQMC, dotted bars).
     Both NNQMC methods are extrapolated according to Ref.\cite{fu2024variance}.
     The shaded region marks the threshold of chemical accuracy (1 kcal/mol).
     Our method consistently achieves or approaches chemical accuracy across the whole benchmark set, aligned well with the $S_{+}$ penalty results.
     It is also more efficient than $S_{+}$ penalty, as it avoids additional penalty terms. 
     {\bf b}. Energy level of the carbon dimer at $r=1.244$ \AA.
     SA-LapNet (blue solid line) provides excitation energies in close agreement with reference methods including semistochastic heat-bath configuration interaction \cite{10.1063/1.4998614} (SHCI, orange dashed line) and natural excited state \cite{NES} (NES, green dashed line). 
     }
     \label{fig:benchmark}
\end{figure*}
In this section, we demonstrate the effectiveness of \nameframework on both ground state and excited states calculations.
We first benchmark the performance of \nameframework in  singlet–triplet energy gap calculations for biradical systems. 
Biradical systems, which contain two unpaired electrons, have the potential in the next-generation organic photovoltaics and molecular magnet~\cite{sing_fission}. The singlet–triplet gap is a fundamental quantity that strongly influences their reactivity and spectroscopic behavior.  
In \cref{fig:benchmark}a, we plot the difference between our calculated results and the experimental references \cite{biradical_lee,biradical_shee} across a diverse set of biradical systems (See \spnote{1} for details).
We also compare with  spin-projected  auxiliary-field quantum Monte Carlo \cite{biradical_lee, biradical_shee} (AFQMC) and the $S_{+}$ penalty-based NNQMC method \cite{spluspenalty}. 
\namenetwork results are in excellent agreement with the experiment values.
Notably, our method does not rely on any penalty term during training.
This represents a significant advance over previous penalty-based approaches in NNQMC for better efficiency and robustness. 

We also study how spin adaptation benefit excited-state calculations by integrating \nameframework with the NES method \cite{NES}.
To showcase its efficacy, we apply it to the carbon dimer, a benchmark system known for its dense, strongly correlated low-lying excited states. We use a high-spin-to-low-spin strategy, as detailed in \cref{subsec: training strategy}, to derive the excited states.
As shown in \cref{fig:benchmark}b, our method provides accurate excitation energies for the low-lying singlet and triplet states of the carbon dimer, consistent with previous NES results \cite{NES} and semistochastic heat-bath configuration
interaction (SHCI) results \cite{10.1063/1.4998614}. 
A key advantage of our spin-adapted method is the separation of the singlet and triplet states, which leads to a reduction of the number of excited states required by each scenario without introducing extra hyperparameters. 
While the penalty-based methods\cite{szabo2024improved,spluspenalty} may also reduce the number of excited states in each scenario, as mentioned before, these methods can introduce additional computational overhead and lead to instable training process under unsuitable hyperparameters.  
More comparisons between penalty method and \nameframework are provided in \spnote{7}.

\FloatBarrier

\subsection{Iron-sulfur Clusters}\label{subsec: iron-sulfur}

\begin{figure*}[t]
    \centering
    \includegraphics[width=0.9\linewidth]{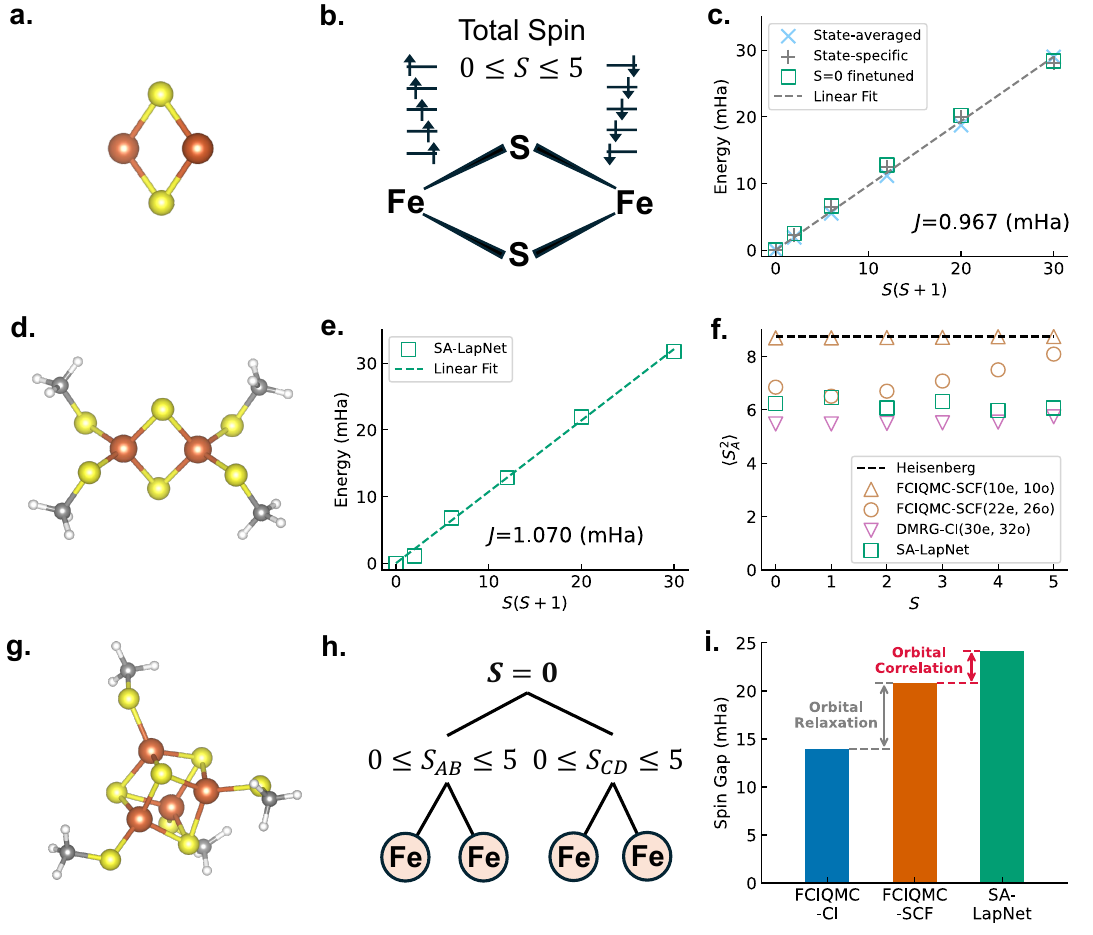}
     \caption{{\bf \namenetwork calculations for Iron-Sulfur Clusters.} In this figure, the balls
of different colors represent different elements: red is iron, yellow is sulfur, gray is carbon, and white is hydrogen.
{\bf a}. Structure of the \ce{[Fe2S2]^{2+}} cluster. 
{\bf b}. Schematic plot of spin coupling for \ce{[Fe2S2]^{2+}}. 
{\bf c}. Energy ladder obtained from state-averaged, state-specific, and $S=0$ finetuned training schemes.
The estimated magnetic coupling constant from state-specific calculation is $J=0.967$ mHa.
{\bf d}. Structure of the \ce{[Fe2S2(SCH3)4]^{2-}} complex. 
{\bf e}. $S=0$ finetuned energy ladder.
The estimated magnetic coupling constant is $J=1.070$ mHa. 
{\bf f}. Local spin $\langle S_A^2 \rangle$ of each state compared with FCIQMC-SCF(10e,10o)\cite{dobrautz2021spin}, FCIQMC-SCF(22e,26o)\cite{dobrautz2021spin}, DMRG-CI(30e,32o)\cite{sharma2014low}, and Heisenberg predictions.
These results indicate that the NNOs can well capture the electron delocalization to sulfurs. 
{\bf g}. Structure of the \ce{Fe4S4(SCH3)4} cluster. 
{\bf h}. Schematic plot of spin-coupling pathways into the $S=0$ state. 
{\bf i}. Spin gap of \ce{Fe4S4(SCH3)4} predicted by \namenetwork  compared with results from FCIQMC-CI and FCIQMC-SCF with active space of (20e,20o) \cite{dobrautz2021spin}. Orbital relaxation refers to the molecular orbital recombination for different spin configurations. The orbital correlation refers to the orbital distortion caused by the many-body correlation. Both orbital relaxation and orbital correlation enlarge the predicted spin gap. }
     \label{fig:iron-sulfur}
\end{figure*}

Iron-sulfur clusters \cite{beinert1997iron} are vital for biological nitrogen fixation, and their unique electronic structure, characterized by strong electron correlation, poses a significant challenge for conventional computational methods like Density Functional Theory (DFT) \cite{noodleman1981valence,noodleman1995orbital,noodleman1992density}, thereby prompting the development of advanced quantum chemistry methods \cite{li2021resolution,sharma2012spin,benediktsson2022analysis,zhai2023multireference}.
In this section, we apply the \namenetwork to study some prototypical iron-sulfur complexes.

We first study the energy ladder of the \ce{[Fe2S2]^{2+}}, the simplest iron-sulfur cluster. As shown in \cref{fig:iron-sulfur}b, the canonical understanding of such an oxidized iron center relies on the analysis of Hubbard model induced by the superexchange mechanism. In this model, the configuration states following the Hund's rule, where d-orbitals are locally parallel coupled on each iron center, forms the non-interacting zeroth-order wavefunctions. 

Within \nameframework framework, this chemical prior knowledge can be naturally integrated with neural network wavefunction by defining the spin part the same as that of the zeroth-order approximation's configuration. We train the neural networks orbital according to a high-spin-to-low-spin scheme, detailed in \cref{subsec: training strategy}. 
Our results are shown in \cref{fig:iron-sulfur}c, including both state-average and state-specific ones. 
Even with only 10 active NNOs, the \namenetwork can provide an accurate anti-ferromagnetic spin spectrum of \ce{[Fe2S2]^{2+}}, escaping the unphysical ferromagnetic state observed in a complete active space (CAS) with 10 one-body orbitals \cite{li2017spin}.

Interestingly, state-average results with shared set of NNOs provide very close energy results  compared with the state-specific ones, which indicates the remarkable similarity among spatial parts of the low-lying eigenstates in the \ce{[Fe2S2]^{2+}}.
This observation is consistent with the minimal Hubbard model description of these systems.
Accordingly, we develop a more efficient training scheme, where the wavefunction with different spin quantum number are finetuned from the NNOs trained for $S=0$ state.
As shown in \cref{fig:iron-sulfur}c, the results from this new scheme, termed as `$S=0$ finetuned' scheme, derive the same energy ladder as the state-specific result. 
Because this pipeline avoids the time-consuming excited state calculation, it enables calculations for even larger systems.

We further conduct calculations on the \ce{[Fe2S2(SCH3)4]^{2-}} system, a model complex with highly strong static and dynamic correlations. 
We compute the energy ladder according to the `$S=0$ finetuned' scheme, with results listed in \cref{fig:iron-sulfur}e. 
To better demonstrate the effectiveness of NNOs, we analyze the local spin of each finetuned state, shown in \cref{fig:iron-sulfur}f. 
Here, we compare the local spin value of \namenetwork with results from advanced quantum chemistry methods under different active spaces, including stochastic-CASSCF via GUGA-FCIQMC as CI solver (FCIQMC-SCF) \cite{dobrautz2021spin}  and density matrix renormalization group (DMRG-CI) \cite{sharma2014low}. 
For a minimal active space of (10e, 10o), the spin remains localized on each iron center. Enlarging the active space to (22e, 26o) results in a decreased spin magnitude on the Fe atom for the low-spin state. However, for high-spin states, the limited number of accessible high-spin configurations causes the spin to remain localized on the Fe atom. Only with the largest active space of (30e, 32o) the spin magnitudes of the Fe atoms decrease uniformly across all states, definitively demonstrating that the electrons delocalize to the sulfur atom. Interestingly, our SA-LapNet, utilizing a compact set of only 10 active NNOs, accurately reproduces these large-active-space results.
This result confirms that the NNOs can effectively capture the electron correlation by distorting the orbitals according to all-electron position. We refer to this phenomenon as the \textit{orbital correlation} effect.

Finally,  we calculate the energy gap between the high-spin state ($S=20$) and the low-spin state ($S=0$) of \ce{Fe4S4(SCH3)4}. As illustrated in \cref{fig:iron-sulfur}h, multiple pathways exist for coupling to the $S=0$ state in this complex four-center Fe-S cluster, presenting a challenge for conventional NNQMC methods in deriving the lowest antiferromagnetic ground state. 
To better describe these spin coupling pathways, we assign 20 NNOs as active orbitals in our calculation. More details are provided in \spnote{6}. 
In \cref{fig:iron-sulfur}i, we compare our predicted spin gaps with FCIQMC-CI(20e,20o) and FCIQMC-SCF(20e,20o) results \cite{dobrautz2021spin}. 

Comparing established methods reveals a clear trend: an improved treatment of electron correlation results in a larger spin gap. Specifically, the FCIQMC-SCF(20e, 20o) method, by introducing orbital relaxation effects, predicts a larger spin gap compared to FCIQMC-CI(20e, 20o). \namenetwork further incorporates the orbital correlation effect, in which the orbitals are dynamically transformed according to all-electron position. As demonstrated by our results on the \ce{[Fe2S2(SCH3)4]^{2-}} system, this ability to capture correlation allows electrons to more effectively delocalize from the iron centers, which in turn reduces the on-site Coulomb repulsion energy and yields a larger, more accurate spin gap for the iron-sulfur system. This underscores that properly accounting for electron correlation is essential for accurate spin-state calculations, and the \nameframework provides a robust and promising route to achieve these high-quality results.

\section{Discussion}\label{sec: Discussion}
We presents a significant advancement in quantum chemistry by introducing \nameframework, effectively addressing spin symmetry in real space.
By explicitly decoupling the wavefunction into its spatial and spin components, we not only solve the spin contamination in NNQMC methods, but also make a major step towards bridging the gap between high-accuracy wavefunction methods and chemically intuitive descriptions.  
The accurate results on the complicated molecular systems demonstrate the effectiveness of \nameframework, bringing new insights in understanding the electronic structure of chemical systems. 

NNQMC methods in real space \cite{hermann2020deep,ferminet} have emerged as a cutting-edge \textit{ab initio} approach, demonstrating gold-standard accuracy in molecular systems \cite{jiang2025neural}. 
Advances in neural network architectures \cite{glehn2023a,li_ab_2022,PhysRevX.14.021030,pescia2024message,li2024computational,scherbela2025accurate,geier2025self,nys2024ab,linteau2025universal} and algorithms \cite{neklyudov2023wasserstein,goldshlager2024kaczmarz} have broadened NNQMC’s applicability, including extensions to excited states \cite{paulinet_excited,NES} and potential energy surface \cite{gao_pesnet_2022,scherbela2024towards,rende2025foundation,foster2025ab}. The \nameframework\ approach can be readily integrated into these developments, providing a rigorous and general treatment of spin symmetry for machine learning-based quantum chemistry methods.

While some prior works \cite{zhan2025expressivity,li2025deep} have used partially spin-space decoupled neural network $\psi_i(\vr_i,\sigma_i|\{\vr\})$ to study systems with non-conserved spin, the connection between the NNO and a spin-adapted ans{\"a}tz has not been previously recognized. Our work formalizes this relationship, providing a robust theoretical foundation for developing spin-adapted neural network ans{\"a}tz. Moreover, even though we consider the spinor-based decomposition of $\Theta_S$, the Hamiltonian studied in this work is still spin-independent. Thus, we use the fixed spin coordinates instead of the continuous spin \cite{PhysRevE.96.043305} in our calculation, following the convention in quantum Monte Carlo \cite{foulkes_quantum_2001}. This fixed spin coordinates can further reduce the number of products to represent $\Theta_S$, as discussed in \spnote{5}. 

From the perspective of theoretical quantum chemistry, our method can be viewed as a natural generalization of the spin-adapted basis \cite{paldus1979configuration,PhysRevLett.27.1105,goddard1973generalized,wang2020describing,dunning2021spin}, where the one-body orbitals are replaced with powerful NNOs. This generalization retains the physical rigor and interpretability of the orbitals while leveraging the power of modern machine learning to achieve higher accuracy. The properties of these NNOs, particularly their relationship to reduced density matrices, present a fascinating avenue for future research.

\section{Methods}\label{sec: Methods}
\newcommand{\Ncore}[0]{N_c}
\subsection{Efficient Implementation of the \nameframework}\label{subsec: implementation}
We develop several methods to enhance the efficiency of \nameframework. The first one is the chemical inspired active orbital selection. Without loss of generality, we assume that the first $2\Ncore\leq N$ orbitals are core orbitals, i.e., $\psi_{2i-1}=\psi_{2i}$ for $i=1,2,...,\Ncore$ and the other $N_a=N-2N_c$ orbitals are active orbitals. Then, we have:
\begin{equation}
    \hat{A}(\Psi\otimes\Theta_S)=\hat{A}\left(\Psi\otimes\Theta^{\text{core}}\Theta^{\text{act}}_{S}\right), \label{eq: closed-shell}
\end{equation}
where %
$\Theta^{\text{act}}_{S}$ is the spin function related to the active orbitals determined by $\Theta_S$, 
$\Theta^{\text{core}}=\chi^\uparrow(\sigma_1)\chi^\downarrow(\sigma_2)...\chi^\uparrow(\sigma_{2\Ncore-1})\chi^\downarrow(\sigma_{2\Ncore})$ is the spin function related to core orbitals. $\chi^{\uparrow/\downarrow}=\frac{1}{2}\pm\sigma$ is the one-body spin-up/spin-down function.
The detailed derivation of \cref{eq: closed-shell} can be found in \spnote{2}. 
As shown in \cref{eq: closed-shell}, the spin functions related to the core orbitals is mathematically equivalent to a direct product of single-body spin functions, thus the number of products $N_{\text{cfg}}$ is determined by the number of active orbitals rather than the number of electrons. The core orbitals are selected after the pretraining of neural network orbitals, in which the orbitals matched to the low energy orbitals from Hartree-Fock calculation are selected as the core orbitals. 

Another benefit of the core orbital selection is that the row related to the core orbitals remains the same across all determinants with the same $\Psi$. This character enables the fast determinant update technique to reduce the computational complexity of determinants. More concretely, the matrix $M_t$ in \cref{eq: sum dets} has the following form:
\begin{equation}
    M^t=\left(\begin{matrix}
        M^t_{ac}&M^t_{aa}\\
        M_{cc}&M_{ca}
    \end{matrix}\right),
\end{equation}
where $M^t_{aa}\in\mathbb{C}^{N_a\times N_a}$, $M^t_{ac}\in\mathbb{C}^{N_a\times 2N_c}$, $M_{ca}\in\mathbb{R}^{2N_c\times N_a}$, $M_{cc}\in\mathbb{R}^{2N_c\times 2 N_c}$ are sub-matrices of $M^t$. Note that $M_{ca}$ and $M_{cc}$ are independent of $t$. Thus, we can leverage a simple determinant identity to enhance the calculation speed:
\begin{equation}
\det(M^t)=\det\left(M^t_{aa}-M^t_{ac}(M_{cc}^{-1}M_{ca})\right). \label{eq: detA-BCD}
\end{equation}
As the calculation results of $M_{cc}^{-1}M_{ca}$ can be shared with each $t$, \cref{eq: detA-BCD} can reduce the computational complexity related to the anti-symmetrization process from $\mathcal{O}(N_{\text{cfg}}N^3)$ to $\mathcal{O}(N_a^3 + N_{\text{cfg}}N^2N_a)$. We implement this identity through the lower-upper decomposition with full pivoting in case of numerical instability. 
The implementation details and corresponding Forward Laplacian \cite{li2024computational} rules are provided in \spnote{3}. 

To enhance the efficiency of \nameframework on polynuclear transition metal systems, we develop an efficient decomposition algorithm based on the Fourier transformation. On these systems, the spin part of wavefunction usually exhibits a locally parallel character. Mathematically speaking, it means that $\Theta_S$ is invariant under the permutation across several spin coordinates \cite{song2024permutation}. Without loss of generality, we assume that the first $\Npara$ spin coordinates are permutation invariant:
\begin{align}
&\Theta_S(\sigma_1,\sigma_2,...,\sigma_{\Npara},...,\sigma_N)\nonumber\\={\ }
&\Theta_S(\sigma_{p(1)},\sigma_{p(2)},...,\sigma_{p({\Npara})},...,\sigma_N),
\end{align}
where 
$p$ is the permutation operation in the symmetric group $\mathcal{S}_{\Npara}$ that permutes the index $i$ to $p(i)$. 

Based on the property of permutation invariant polynomial, there is an efficient partially sum of products decomposition for $\Theta_S$ with Fourier transformation:
\begin{equation}
\label{eq: ft decompose}
    \Theta_S=\sum_{i=0}^{\Npara-1}\prod_{j=1}^{\Npara}\chi_{i}(\sigma_j)\tilde{\Theta}^i(\sigma_{\Npara+1:N}),
\end{equation}
where $\chi_{i}=\frac{1}{Q_{i}}\left(\chi^{\uparrow}+\exp\left(\mbi\frac{2\pi}{\Npara}i\right)\chi^{\downarrow}\right)$. $\mbi$ is the imaginary unit. $Q_i$ is the normalizing constant. 
$\tilde{\Theta}^i$ is the remained spin function that can be derived from Fourier transformation. %
See \spnote{4} for details. 
Here, the number of products used in \cref{eq: ft decompose} is $\Npara$, significantly smaller than that of decomposing it with $\chi^{\uparrow}$ and $\chi^{\downarrow}$, which requires $\mathcal{O}(2^{N_{\text{cfg}}})$ terms \cite{marti2025spin}.
Applying this method to all the locally parallel coupled orbitals can
significantly reduce $N_{\text{cfg}}$.

\subsection{Wavefunction and Optimization}\label{subsec: optimization}
In this paper, we modify the LapNet~\cite{li2024computational} with \nameframework to represent the spin-adapted wavefunction. LapNet is an attention-based ansatz in continuous space, which takes the spatial and spin coordinates of electrons as input and processes them through sparse derivative attention blocks. To satisfy the constraints imposed by the \nameframework, these spin coordinates are excluded from the input representation of the neural network, i.e, the spatial part. The pseudo-code of computing the wavefunction is provided in \cref{alg:SA-LapNet}. Here, we adopt the Jastrow factor used in Ref.\cite{huang1998spin} to reduce the numerical instability caused by the cusp condition. 
\renewcommand{\baselinestretch}{1.5}
\begin{algorithm*}
\caption{\namenetwork}\label{alg:SA-LapNet}
\begin{algorithmic}[1]
\Require Neural network parameters $\theta$, electron positions $\vr_1, \vr_2,...,\vr_N$, nuclei positions $\mR_1,\mR_2,...,\mR_{N_{A}}$, 
the spin part related to active orbitals $\Theta_{S}^{\text{act}}$
\State $\vg_i^0=\vf_i^0 \gets \text{concat}\left(\left\{\frac{\vr_i-\mR_I}{\|\vr_i-\mR_I\|}\ln\|\vr_i-\mR_I\|,\ln\|\vr_i-\mR_I\|, I=1,2,...,N_A\right\}\right)$
\For{$l=0,1,2,...,L-1$}
\State $\tilde{\vf}^l\gets\text{Attn}(\vg^l,\vg^l,\vf^l)$
\State $\vf^{l+1}\gets\text{MLP}(\tilde{\vf}^l)$, $\vg^{l+1}\gets\text{MLP}(\vg^l)$
\EndFor
\State $\vo_i\gets\text{env}(\vr_i)\odot\vf_i^{L}$ 
\State $\vo_i^{\text{core}}\gets[\vo_i]_{1:\Ncore}$,  $\vo_i^{\text{act}}\gets[\vo_i]_{N_c+1:N_c+N_a}$
\State $M_c^{\uparrow}\in \mathbb{R}^{N_c\times N^{\uparrow}}\gets\text{concat}(\vo_i^{\text{core}}, i\leq N^{\uparrow})$
\State $M_c^{\downarrow}\in \mathbb{R}^{N_c\times N^{\downarrow}}\gets\text{concat}(\vo_i^{\text{core}},N^{\uparrow}<i\leq N)$
\State $M_c\in\mathbb{R}^{2N_c\times N}\gets\left(\begin{matrix}
    M_{c}^{\uparrow}&0\\
    0& M_{c}^{\downarrow}
\end{matrix}\right)$
\State$M_a^{\uparrow}\in \mathbb{R}^{N_a\times N^{\uparrow}}\gets\text{concat}(\vo_i^{\text{act}}, i\leq N^{\uparrow})$
\State $M_a^{\downarrow}\in \mathbb{R}^{N_a\times N^{\downarrow}}\gets\text{concat}(\vo_i^{\text{act}},N^{\uparrow}< i\leq N)$
\State decompose $\Theta^{\text{act}}_{S}=\sum_{t=1}^{N_{\text{cfg}}}c_t\chi_1^t\chi_2^t...\chi_{N_a}^t$, $\chi_i^{t}=\alpha_i^t\chi^{\uparrow} + \beta_i^t\chi^{\downarrow}$
\State $\boldsymbol{\alpha}^t\gets\text{concat}(\alpha_i^t)$, $\boldsymbol{\beta}^t\gets\text{concat}(\beta_i^t)$, 

\For{$t=1,2,...,N_{\text{cfg}}$}
\State $M^t\in\mathbb{C}^{N\times N}\gets\left(\begin{matrix}
    M_a^{\uparrow}\text{Diag}(\boldsymbol{\alpha}^t),M_a^{\downarrow}\text{Diag}(\boldsymbol{\beta}^t)\\M_c
\end{matrix}\right)$
\EndFor
\State $J\gets \exp\left(-\sum_{i<j}\frac{1}{2}\frac{1}{1+\|\vr_i-\vr_j\|}\right)$
\State $\Phi\gets J\sum_{t=1}^{N_{\text{cfg}}}c_t\text{Re}\left[\text{det}[M^t]\right]$\\
\Return $\Phi$
\end{algorithmic}
\end{algorithm*}
\renewcommand{\baselinestretch}{1}

For the molecular and atomic systems studied in \cref{subsec:biradical}, we adopt a two-stage training protocol analogous to that used in FermiNet \cite{ferminet}. In the first stage, i.e., the pretraining stage, we train the NNOs to match the orbitals derived from the Hartree-Fock method. In the second stage, the total energy of the wavefunction serves as the loss function $E_\theta$:
\begin{equation}
    E_\theta=\frac{\bra{\Psi_{\theta}}{\hat{H}}\ket{\Psi_{\theta}}}{\braket{\Psi_{\theta}}{\Psi_{\theta}}}=\mathbb{E}_{\vr\sim p_{{\theta}}}[E_L],\label{eq: total energy}
\end{equation}
where $\theta$ denotes the parameters of the neural networks. $\Psi_\theta$ is the neural network wavefunction. $p_{\theta}=\frac{|\Psi_{\theta}|^2}{\int |\Psi_{\theta}|^2}$ is the normalized probability distribution corresponding to $|\Psi_{\theta}|^2$. $\hat{H}$ is the Hamiltonian of the system. $E_{L}=\hat{H}\Psi_{\theta}/{\Psi_{\theta}}$ is the local energy function. We calculate the total energy $E_{\theta}$ and the corresponding gradient $\partial_{\theta}E_\theta=2\mathbb{E}_{\vr\sim p_{{\theta}}}[(E_L-E_{\theta})\partial_{\theta}\ln|\Psi_{\theta}|]$ with the Monte Carlo method, where the walkers are generated by the Metropolis adjusted Markov Chain Monte Carlo method with Gaussian proposal. 

We use the NES method to calculate high excited states in this work. The NES method was originally introduced by \citet{NES}, where the problem of computing the first $K$ excited states is reformulated as a variational energy minimization problem defined in an extended Hilbert space. 
This extended Hilbert space is the direct product of $K$ conventional Hilbert spaces $\mathcal{H}:\mathbb{R}^{3N}\rightarrow\mathbb{R}$ that correspond to the $N$ electrons. More concretely, the energy of the first $K$ eigenstate in an electron system can be derived from the following minimization problem:
\newcommand{\vrall}[0]{\vr^{\text{all}}}
\begin{equation}
    \min_{\theta} \mathbb{E}_{(\vrall_{1},\vrall_{2},...,\vrall_{K})\sim \det({\Psi_{\text{mat}}})^2}\text{Tr}[(\hat{H}\Psi_{\text{mat}})\Psi_{\text{mat}}^{-1}], \label{eq: NES objective}
\end{equation}
where $\Psi_{\text{mat}}\in\mathbb{R}^{K\times K}$ is defined elementwise as:
\begin{equation}
    [\Psi_{\text{mat}}]_{ij}=\Psi_{i,\theta}(\vrall_j),{\ }i,j=1,...,K.
\end{equation}
 and each $\Psi_{i,\theta}:\mathbb{R}^{3N}\rightarrow \mathbb{R}$ is a state represented by the neural network. $\vrall_i\in\mathbb{R}^{3N}$ is the spatial coordinates of all electrons on the $i$-th conventional Hilbert space. After optimization, the energy level of the first $K$ states is computed through diagonalizing the following matrix:
\begin{equation}
F_{\Psi_{\text{mat}}}=\mathbb{E}_{(\vrall_{1},\vrall_{2},...,\vrall_{K})\sim \det({\Psi_{\text{mat}}})^2}(\hat{H}\Psi_{\text{mat}})\Psi_{\text{mat}}^{-1}. 
\end{equation}

\subsection{Training Strategies of NNO }
\label{subsec: training strategy}
In the systems that have strong static correlation, 
the orbitals derived from Hartree-Fock method are qualitatively wrong for the low-spin state, making the Hartree-Fock pretraining stage described in \cref{subsec: optimization} suboptimal. To address this limitation, we develop a high-spin–to–low-spin training strategy tailored for these systems. 
In the first step of this strategy, i.e., the high-spin training step, we train the NNOs on a high-spin state, in which the static correlation is significantly reduced due to the spin constraint.
In the second step of this strategy, i.e., the low-spin training step, the wavefunctions for low-spin states are initialized by the NNOs obtained from the first step and are further optimized by the standard VMC method for the ground state or the NES method for excited states. 

On the \ce{[Fe2S2]^{2+}} system, we observe that optimizing different spin states in a state-averaged way at first can provide better energy for intermediate spin states. A subsequent independent optimization of each state, i.e., state-specific training,  is applied to derive the lowest energy for each spin quantum number. However, the state-averaged training is so expensive for the large \ce{[Fe2S2(SCH3)4]^{2-}} and \ce{Fe4S4(SCH3)4} system. Inspired by the transferability of the NNOs trained on $S=0$ state, for these large systems, we first train the $S=0$ state with a high-spin-to-low-spin strategy. Then, we initialize the other spin states with the orbitals trained from $S=0$ state and optimize these states independently to derive the energy with different spin quantum number.

\backmatter

\bmhead{Data Availability}

All data supporting the findings of this study are available within the Supplementary Information.

\bmhead{Acknowledgments}

We thank William A. Goddard III, Giovanni Li Manni, Zhendong Li and Qiming Sun for the insightful discussions.
We thank the ByteDance Seed Group for inspiration and encouragement. 
We also thank Hang Li for his guidance and support.
Ji Chen is supported by the National Key R\&D Program of China (2021YFA1400500) and National Science Foundation of China (12334003). Liwei Wang is supported by National Science and Technology Major Project (2022ZD0114902) and National Science Foundation of China (NSFC92470123, NSFC62276005). Di He is supported by National Science Foundation of China (NSFC62376007).

\bmhead{Competing Interests}

The authors declare no competing interests.

\bibliography{sn-bibliography}
\end{document}


\tableofcontents
\section{Hyperparameters and System Configurations}
The default hyperparamters used in this study are listed in \cref{table:default params}. For all the ground state optimization problems, we use the single-precision floating-point. For all the excited state optimization, we use the double-precision floating-point for numerical stability. 

For the biradical systems, we use the geometries from Ref.\cite{spluspenalty}. The energies of the singlet state and triplet state are extrapolated according to the method proposed in Ref.\cite{fu2024variance}, which is aligned with original calculation setup in Ref.\cite{spluspenalty}. We usually assign 3 pairs of orbitals that couple to open-shell singlet for both the singlet and triplet system, except the smallest carbon atomic system where we only assign 2 pairs of open-shell coupled orbitals. A pair of parallel coupled orbitals is additionally assigned for the triplet systems. 

For the iron-sulfer clusters, we use the local psuedopotential on the Fe atom and S atom to accelerate the calculation \cite{PH}. The initial learning rates on these systems are 0.025. For the \ce{[Fe2S2]^{2+}} cluster, we use the geometry from Ref.\cite{benediktsson2022analysis}. For the \ce{[Fe2S2(SCH3)4]^{2-}} complex, we use the geometry from Ref.\cite{sharma2014low}.  For the \ce{Fe4S4(SCH3)4}, we use the geometry from Ref.\cite{dobrautz2021spin}.  We use the high-spin-to-low-spin strategy to derive the energy ladder of these clusters. For \ce{[Fe2S2]^{2+}} and \ce{[Fe2S2(SCH3)4]^{2-}}, we train neural networks for 100,000 steps in each stage. For the \ce{Fe4S4(SCH3)4},  we train neural networks for 200,000 steps in each stage.

\begin{table*}[htbp]
    \centering
    \caption{Default hyperparameters.}
    \begin{tabular}{ccc}
    \toprule\midrule
        & Parameter & Value \\
    \midrule
    \multirow{6}{*}{Training}& 
    Optimizer & KFAC\cite{10.5555/3045118.3045374} \\
    & Iterations & 2e5 \\
    & Batch size & 4096 \\
    & Learning rate $\eta$ at iteration $t$ & $\eta_0/(1+\frac{t}{t_{\mathrm{delay}}})$ \\
    & Learning rate decay $t_{\mathrm{delay}}$ & 1e4\\
    & Initial learning rate $\eta_0$ & $0.05$ \\
    & Local energy clipping & 5.0 \\
    \midrule
    \multirow{4}{*}{Pretraining} & Optimizer & LAMB\cite{You2020Large} \\
    & Iterations & 5e3 or 2e4 \\
    & Basis set & aug-cc-pVDZ \\
    & Learning rate &3e-4\\
    \midrule
    \multirow{3}{*}{MCMC} & Decorrelation steps & 30 \\
    & Proposal standard deviation & 0.02 \\
    & Blocks & 1 \\
    \midrule
    \multirow{4}{*}{KFAC} & Norm constraint & $1\text{e-3}$ \\
    & Damping & 1e-3 \\
    & Momentum & 0 \\
    & Covariance moving average decay & 0.95 \\
    \midrule\bottomrule
    \end{tabular}
    \vspace{-0.3cm}
    \label{table:default params}
\end{table*}
\section{Core Orbitals in \nameframework Framework}

In this note, we will prove that as long as an orbital $\psi(\cdot|\{\vr\})$ appear twice in the spatial part $\Psi$, then the related spin part can only be singlet. The spin part can be further represented by a product of one-body spin functions. Without loss of generality, consider the situation where the two electrons occupy the same spatial orbital, we have
\begin{equation}
    \Psi(\vr_1,\vr_2,\vr_3,...,\vr_N)=\tilde{\psi}(\vr_1|\{\vr\})\tilde{\psi}(\vr_2|\{\vr\})\psi_3(\vr_3|\{\vr\}) ... \psi_N(\vr_N|\{\vr\}). \label{eq: permute inv}
\end{equation}
\newcommand{\Pspaceij}[1]{\mathcal{P}_{#1}^{\text{space}}}
\newcommand{\Pspinij}[1]{\mathcal{P}_{#1}^{\text{spin}}}
\cref{eq: permute inv} implies that $\Psi$ is invariant under the permutation of the first 2 electron's spatial coordinate. More concretely, let $\Pspaceij{ij}$ denote the exchange operator of spatial coordinate $i$ and $j$ and $\Pspinij{ij}$ denote the exchange operator of spin coordinate $i$ and $j$, we have $\Pspaceij{12}\Psi=\Psi$. Then, leveraging the property of anti-symmetrization operator $\hat{A}$, we have: 
\begin{equation}
    \hat{A}(\Psi\otimes\Theta_S) =\hat{A}\left(\frac{1}{2}(1+\Pspaceij{12})\Psi\otimes\Theta_S\right) =\frac{1}{2}\left(\hat{A}\left(\Psi\otimes\Theta_S\right) - \hat{A}\left(\Psi\otimes\Pspinij{12}\Theta_S\right)\right)=\hat{A}\left(\Psi\otimes\frac{1}{2}(1-\Pspinij{12})\Theta_S\right).
\end{equation}
This relation reveals that for the total wavefunction to be antisymmetric, the spin function $\Theta_S$ must have a component that is antisymmetric with respect to the exchange of the spins of the two electrons in the doubly occupied orbital. This antisymmetric spin component is proportional to the singlet state, $f^{\text{anti}}(\sigma_1,\sigma_2)=\sigma_1-\sigma_2$, which is the only two-electron spin function that is antisymmetric under particle exchange:
\begin{equation}
    (1-\Pspinij{12})\Theta_S = f^{\text{anti}}(\sigma_1,\sigma_2)\tilde\Theta_S(\sigma_3,\sigma_4,...,\sigma_N), 
\end{equation}
where $\tilde\Theta_S$ is a spin function independent to the first 2 spin coordinates.
Moreover, notices that the singlet spin function $f^{\text{anti}}=\chi^\uparrow(\sigma_1)\chi^\downarrow(\sigma_2)-\chi^\downarrow(\sigma_1)\chi^\uparrow(\sigma_2)=(1-\Pspinij{12})\chi^\uparrow(\sigma_1)\chi^\downarrow(\sigma_2)$, then we have:
\begin{equation}
        \hat{A}(\Psi\otimes\Theta_S)=\hat{A}\left(\Psi\otimes\frac{1}{2}(1-\Pspinij{12})\chi^\uparrow(\sigma_1)\chi^\downarrow(\sigma_2)\tilde\Theta_S\right)=\hat{A}\left(\frac{1}{2}(1+\Pspaceij{12})\Psi\otimes\chi^\uparrow(\sigma_1)\chi^\downarrow(\sigma_2)\tilde\Theta_S\right)=\hat{A}\left(\Psi\otimes\chi^\uparrow(\sigma_1)\chi^\downarrow(\sigma_2)\tilde\Theta_S\right). \label{eq: closed-shell3}
\end{equation}
With \cref{eq: closed-shell3}, we shows that the double occupied spatial orbitals will lead to a direct product decomposition of the corresponding spin coordinates. Applying this decomposition to all the core orbitals, we can prove the main text Eq.(5).

\section{Numerical Stable Fast Determinants Update}
In this note, we demonstrate how to efficiently and accurately compute the determinant of 
\begin{equation}
    M^t=\left(\begin{matrix}
        M^t_{ac}&M^t_{aa}\\
        M_{cc}&M_{ca}
    \end{matrix}\right),
\end{equation}
over different configurations. Here, $M^t\in \mathbb{C}^{N\times N}$, $M^t_{aa}\in\mathbb{C}^{N_a\times N_a}$, $M^t_{ac}\in\mathbb{C}^{N_a\times 2N_c}$, $M_{ca}\in\mathbb{R}^{2N_c\times N_a}$, $M_{cc}\in\mathbb{R}^{2N_c\times 2 N_c}$ are sub-matrices of $M^t$. For easy of reference, we define $M_c=(M_{cc}, M_{ca})\in \mathbb{C}^{2N_c\times N}$ and $M_a^t=(M_{ac}, M_{aa})\in \mathbb{C}^{N_a\times N}$.
We first decompose $M_c$ with full-pivoting LU decomposition:
\begin{equation}
    M_c = PLUQ,
\end{equation}
where $P\in \mathbb{N}^{2N_c\times2N_c}$ and $Q\in \mathbb{N}^{N\times N}$ are row and column permutation matrix; $L\in \mathbb{C}^{2N_c\times 2N_c}$ is a lower triangular matrix with unit diagonal terms; $U\in \mathbb{C}^{2N_c\times N}$ is an upper triangular matrix. Then, with the Schur complement, we have:
\begin{equation}
    \det(M^t)=\det(P)\det(Q)\det(U_{cc})\det(\tilde {M}^t_{aa}-(U_{ac}U_{cc}^{-1})\tilde{M}^t_{ca}),
\end{equation}
where $U_{cc}=U_{:,:2N_c}$ and $U_{ca}=U_{:,2N_c:}$ are submatrices of $U$; $\tilde {M}_a^t=M_a^t Q^{-1}$ is the permuted $M_a^t$; $\tilde {M}_{ac}^t=[\tilde {M}_a^t]_{:,:2N_c}$ and $\tilde {M}^t_{aa}=[\tilde {M}_a^t]_{:,2N_c:}$ are submatrices of $M_a^t$. Here, as $U_{cc}$ and $U_{ac}U_{cc}^{-1}$ are independent of $t$, we can share these computational results over different configurations, thereby saving the computation. 

To apply the forward Laplacian method, we can define the derivative and Laplacian of $U$ according to the following equations:
\begin{equation}
    \nabla M_c=PL\nabla U Q, \quad \nabla^2 M_c=PL\nabla^2 U Q.
    \label{eq: U FL}
\end{equation}
As $L$, $Q$, $P$ are invertible matrices, the derivative and Laplacian of $U$  can be derived from \cref{eq: U FL}. Thus, we can directly apply the forward Laplacian method on the computation graph of Schur complement to compute the derivative and Laplacian of $\det(M^t)$. 

\section{Fourier Transformation-based Decomposition}

In this note, we focus on deriving an efficient decomposition of $\Theta_S$ with partially permutation symmetry. Without loss of generality, we assume that $\Theta_S$ is invariant under the permutation over the first $\Npara$ spin coordinates:
\begin{equation}
\Theta_S(\sigma_1,\sigma_2,...,\sigma_{\Npara},...,\sigma_N)\nonumber=\Theta_S(\sigma_{p(1)},\sigma_{p(2)},...,\sigma_{p({\Npara})},...,\sigma_N),
\end{equation}
where $p$ is the permutation operation in the symmetric group $\mathcal{S}_{\Npara}$ that permutes the index $i$ to $p(i)$. 

As these spins are locally parallel couple, the spin function $\Theta_S$ is also the eigen function of the $\hat{S}_{\text{loc}}^2=\left(\sum_{i=1}^{\Npara} \hat{s}_i\right)^2$ with eigenvalue $\frac{\Npara}{2}\left(\frac{\Npara}{2}+1\right)$. Thus, it can be decomposed according to the spin eigenfunctions of the first $\Npara$ electrons:

\begin{equation}
    \Theta_S=\sum_{s_z=-\frac{\Npara}{2}}^{-\frac{\Npara}{2}}h^{\frac{\Npara}{2}, s_z}_{\Npara}\tilde{\Theta}^{s_z}_S,\label{eq: eigen decompose}
\end{equation}
where $h^{s,s_z}_{\Npara}$ is the normalized spin eigenfunction of $\Npara$ electrons with spin quantum number $s$ and $z$-axis spin quantum number $s_z$, $\tilde{\Theta}^{s_z}_S$ is the partial inner product between $\Theta_S$ and $h^{\frac{\Npara}{2}, s_z}_{\Npara}$. To rewrite \cref{eq: eigen decompose} as a sum over product decomposition, we consider the generator function of $h^{\frac{\Npara}{2}, s_z}_{\Npara}$:
\begin{equation}
    g(\lambda|\{\sigma_i\}_{i=1}^{\Npara})=\prod_{i=1}^{\Npara}\left(\chi^{\uparrow}(\sigma_i)+\lambda\chi^{\downarrow}(\sigma_i)\right)=\sum_{i=0}^{\Npara}\lambda^iG_{\Npara,i}h^{\Npara/2,\Npara/2-i}_{\Npara}.\label{eq: generator functions}
\end{equation}
Here, we define 
\begin{equation}
    G_{N,k}=\sqrt\frac{N!}{k!(N-k)!}\label{eq:sqr binorminal}
\end{equation}
as the square root of the binomial coefficients to simplify the expression. $g(\lambda|\{\sigma_i\}_{i=1}^{\Npara})$ is a polynomial of $\lambda$, which is also the product of one-body spin functions $\chi^{\uparrow}+\lambda\chi^\downarrow$. The \cref{eq: generator functions} reveals that the spin eigenfunction $h_{\Npara}^{\Npara/2,s_z}$ is the coefficients of $g(\lambda|\{\sigma_i\}_{i=1}^{\Npara})$. Thus, we can use Fourier transformation to represents the coefficients with the value of polynomial:

\begin{equation}
    h_{\Npara}^{\Npara/2,\Npara/2-i} = \frac{1}{\Npara G_{\Npara,i}}\sum_{j=0}^{\Npara-1}(\bar\xi_{\Npara})^{ij}g((\xi_{\Npara})^j|{\{\sigma_i\}_{i=1}^{\Npara}}), 
    \label{eq: fourier transformation}
\end{equation}
where $\xi_{N}=\exp\left({\mbb\frac{2\pi}{N}}\right)$ is the root of unity. Substituting \cref{eq: fourier transformation} into \cref{eq: eigen decompose}, we can derive the main text Eq.(9).

\section{Spin Functions Represented by the Binary Tree}
\begin{figure}
    \centering
    \includegraphics[width=0.6\linewidth]{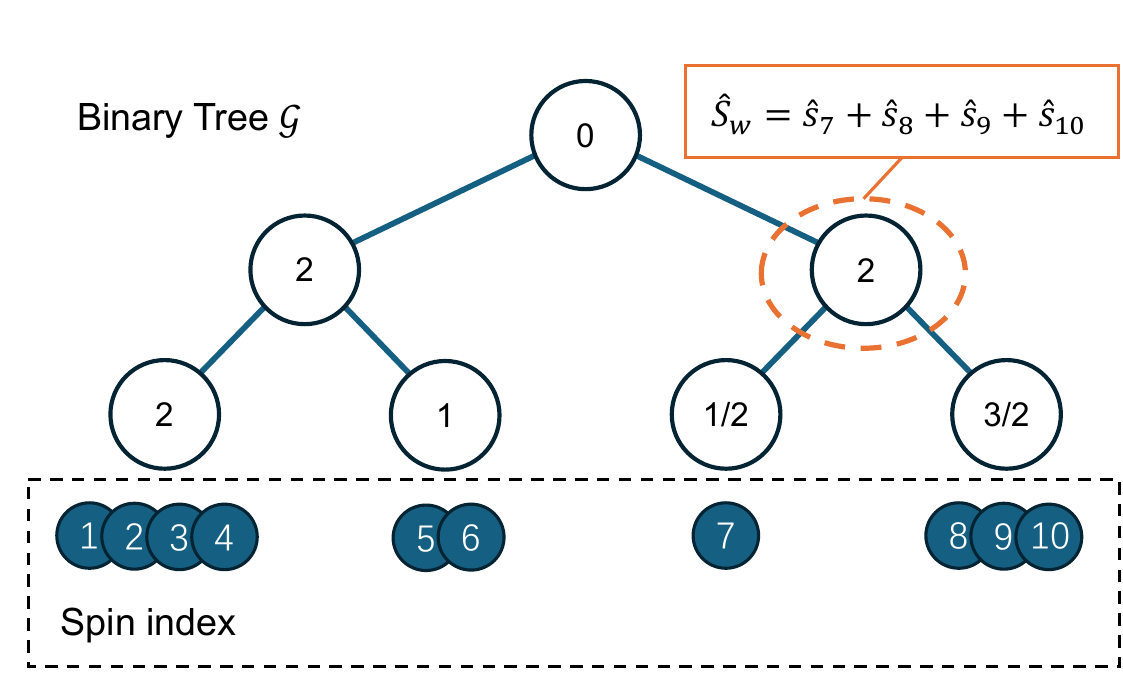}
    \caption{{\bf An example of binary tree representation of spin functions}. The white circles represent to the node on the binary tree $\gG$. The blue circles represent the spin indexes. The number on the node of $\gG$ represents the total spin quantum number of all the spin indexes that belongs to it. For example, spin indexes $7$ to $10$ belong to the node highlighted by the yellow boxes. Thus, the number on the node represents the spin quantum number related to the operator $\hat{S}_w=\hat{s}_7+\hat{s}_8+\hat{s}_9+\hat{s}_{10}$}
    \label{fig:binary tree}
\end{figure}
To construct a spin function in the spin space, we can leverage the 2-body Clebsch–Gordan(CG) coefficients of $SU(2)$ group. We use a binary tree $\gG$ to represents the coupling in these functions. Here, we define a surjection mapping $f_{\text{node}}:[1,2,...,N]\rightarrow\text{Leaf}(\gG)$ that maps the electron index to the leaf node of the binary tree. In another word, each electron index is assigned to one leaf node of $\gG$. The correspond mapping $f_{\text{id}}(w)=\{i|f_{\text{node}}(i)=w\}$ determinants the electron indexes that are assigned to the leaf node $w$. Furthermore, each node $w\in\gG$ is assigned with an integer/half-integer $S_{w}$ that represents the total spin quantum number of electron that belongs to this node.  Here, we say that an electron $i$ belongs to a node $w$ as long as $f_{\text{id}}(i)$ is the descendant of node $w$. $S_{w}$ should satisfy three constraints: 
\begin{enumerate}
    \item For the root node $w_{\text{rt}}$, $S_{w_{\text{rt}}}=S$.
    \item $\forall w\in \text{Leaf}(\gG)$, $S_w= |f_{\text{id}}(w)|/2$.
    \item $\forall w\in \text{NonLeaf}(\gG)$, $|S_{l(w)}-S_{r(w)}|\leq S_{w}\leq S_{l(w)}+S_{r(w)}$.
\end{enumerate}
Here, $\text{NonLeaf}(\gG)$ denotes the set of internal nodes of $\gG$. $l(w)$ and $r(w)$ represent the left child and right child of node $w$, respectively. The expression of the corresponding spin functions is:
\newcommand{\SpinFunc}[0]{\Theta_{S}(\sigma_1,\sigma_2,...\sigma_N|\gG)}
\newcommand{\NLeaf}[0]{\text{NonLeaf}(\gG)}
\newcommand{\Leaf}[0]{\text{Leaf}(\gG)}
\newcommand{\LeafEigen}[0]{h_{2S_{a_t}}^{S_{a_t},m_{a_t}}(\{\sigma_i|i\in f_\text{id}(a_t)\})}
\newcommand{\Binternal}[0]{B^{S_wm_w}_{S_{l(w)}m_{l(w)},S_{r(w)}m_{r(w)}}}
\newcommand{\Btildeinternal}[0]{\tilde{B}^{S_wm_w}_{S_{l(w)}m_{l(w)},S_{r(w)}m_{r(w)}}}
\newcommand{\RootNode}[0]{w_{\text{rt}}}

\begin{equation}
    \SpinFunc=\sum_{m_v|v\in{\gG}, v\neq \RootNode}\prod_{w\in \NLeaf}\Binternal\prod_{t=1}^o\LeafEigen, \label{eq: CG}
\end{equation}
where $B^{S_1m_1}_{S_{2}m_{2},S_{3}m_{3}}$ is the two-body CG coefficients of $SU(2)$ group. $a_t\in \Leaf$ are the ordered leaf nodes of $\gG$, $o=|\Leaf|$. For the non-root node $v\in \gG$, the dummy index $m_v$ ranges from $-S_v$ to $S_v$. For the root node $\RootNode$, the index $m_{\RootNode}$ is decided by the number of spin-up and spin-down electron, i.e., $m_{\RootNode}=\frac{\Nup-\Ndown}{2}$. An example of this representation is shown in \cref{fig:binary tree}. With this binary tree representation, the spin function is represented by the sum over products of the spin eigen functions $\LeafEigen$, whose one-body products decomposition can be efficiently finished by the Fourier transformation. For clearity, we first rewrite $\SpinFunc$ as:
\begin{align}
    &\SpinFunc\\
    =&
    \sum_{m_v|v\in{\gG}, v\neq \RootNode}\prod_{w\in \NLeaf}\Binternal\prod_{t=1}^o\LeafEigen\\
    =&\sum_{m_v|v\in{\gG}, v\neq \RootNode}\prod_{w\in \NLeaf}\Binternal\prod_{t=1}^o\frac{1}{G_{\Npara,S_{a_t}-m_{a_t}}}G_{2S_{a_t},S_{a_t}-m_{a_t}}\LeafEigen\\
    =&\frac{1}{G_{2S,S-m_{\RootNode}}}\sum_{m_v|v\in{\gG}, v\neq \RootNode}\prod_{w\in \NLeaf}\Btildeinternal\prod_{t=1}^oG_{2S_{a_t},S_{a_t}-m_{a_t}}\LeafEigen.\label{eq: spin func full}
\end{align}
Here, we define normalized two-body CG coefficients 
\begin{equation}
    \tilde{B}^{S_1m_1}_{S_2m_2,S_3m_3}=B^{S_1m_1}_{S_2m_2,S_3m_3}\frac{G_{2S_1,S_1-m_1}}{G_{2S_2,S_2-m_2}G_{2S_3,S_3-m_3}},
\end{equation}
to simplify the following derivation. The $G_{N,k}$ is defined by Eq.\ref{eq:sqr binorminal}. We then the define tensor $F^{\gG}$ whose element are given by:
\newcommand{\tFG}[0]{F^{\gG}_{m_{a_1},m_{a_2},...,m_{a_o}}}
\begin{align}
    \tFG=\frac{1}{G_{2S,S-m_{\RootNode}}}\sum_{m_v|v\in{\NLeaf}, v\neq \RootNode}\prod_{w\in \NLeaf}\Btildeinternal.
\end{align}
Intuitively speaking, tensor $F^{\gG}$ is derived from the tensor contraction of normalized CG coefficients, where all the dummy variable $m_v$ related to internal nodes of $\gG$ are summed. We can also define the remained part in \cref{eq: spin func full} as another tensor depends on $\sigma_i$:
\newcommand{\tEsigma}[0]{E^{\sigma_1\sigma_2...\sigma_N}_{m_{a_1},m_{a_2},...,m_{a_o}}}
\begin{align}
\tEsigma=\prod_{t=1}^oG_{2S_{a_t},S_{a_t}-m_{a_t}}\LeafEigen\label{eq: tesigma}
\end{align}
The $\SpinFunc$ can thus be rewritten as the tensor inner product between $F^{\gG}$ with $E^{\sigma_1\sigma_2...\sigma_N}$:
\begin{equation}
    \SpinFunc=\sum_{m_{a_t}}\tFG\tEsigma
\end{equation}
We then insert an identity operator in the tensor contraction under the Fourier basis:
\newcommand{\xiSaMa}[1]{\frac{\xi_{2S_{a_{#1}}}^{(S_{a_{#1}}-m_{a_{#1}})j_{#1}}}{\sqrt{2S_{a_{#1}}}}}
\newcommand{\xiMaSa}[1]{\frac{\xi_{2S_{a_{#1}}}^{(m'_{a_{#1}}-S_{a_{#1}})j_{#1}}}{\sqrt{2S_{a_{#1}}}}}
\newcommand{\ftFG}[0]{\tilde{F}^{\gG}_{j_1j_2...j_o}}
\newcommand{\ftEsig}[0]{\tilde{E}^{\sigma_1\sigma_2...\sigma_N}_{j_1j_2...j_o}}
\begin{align}
    &\SpinFunc\\
    =&\sum_{m_{a_t}}\tFG\tEsigma\\
=&\sum_{m'_{a_t},m_{a_t}}F^{\gG}_{m'_{a_1},m'_{a_2},...,m'_{a_o}}\left(\prod_{t=1}^o\delta_{m'_{a_t},m_{a_t}}\right)\tEsigma\label{eq: insert identity}\\
    =&\sum_{m'_{a_t},m_{a_t}}F^{\gG}_{m'_{a_1},m'_{a_2},...,m'_{a_o}}\left(\prod_{t=1}^o\sum_{j_t=0}^{2S_{a_t}}\frac{\xi_{2S_{a_t}}^{(m'_{a_t}-S_{a_t}+S_{a_t}-m_{a_t})j_t}}{2S_{a_t}}\right)\tEsigma\\
=&\sum_{j_t}\left(\sum_{m'_{a_t}}F^{\gG}_{m'_{a_1},m'_{a_2},...,m'_{a_o}}\xiMaSa{1}\xiMaSa{2}...\xiMaSa{o}\right)\left(\sum_{m_{a_t}}\tEsigma\xiSaMa{1}\xiSaMa{2}...\xiSaMa{o}\right)\\
=&\sum_{j_t}\text{conj}({\ftFG})\ftEsig. \label{eq: ft product}
\end{align}
Here, we define:
\begin{align}
\ftFG&=\sum_{m_{a_t}}F^{\gG}_{m_{a_1},m_{a_2},...,m_{a_o}}\xiSaMa{1}\xiSaMa{2}...\xiSaMa{o},\\
    \ftEsig&=\sum_{m_{a_t}}\tEsigma\xiSaMa{1}\xiSaMa{2}...\xiSaMa{o},\label{eq: ftesig}
\end{align}
as the Fourier transformed $F^{\gG}$ and $E^{\sigma_1\sigma_2...\sigma_N}$. Here, $\ftFG\in\mathbb{C}$ are some constant that can be computed before the neural network training. Leveraging \cref{eq: tesigma} and \cref{eq: generator functions}, $\ftEsig$ can be rewritten as a product of one-body spin functions:
\begin{equation}
    \ftEsig=\prod_{t=1}^o\frac{1}{\sqrt{2S_{a_t}}}g(\xi_{S_{a_t}}^{j_t}|\{\sigma_i|i\in f_{\text{id}}(a_t)\})=\prod_{t=1}^o\frac{1}{\sqrt{2S_{a_t}}}\prod_{i\in f_{\text{id}}(a_t)}(\chi^\uparrow(\sigma_i)+\xi_{S_{a_t}}^{j_t}\chi^{\downarrow}(\sigma_i)). \label{eq: E functions}
\end{equation}
Substituting \cref{eq: E functions} into \cref{eq: ft product}, we then find a sum-over-product decomposition of $\SpinFunc$ through Fourier transformation. Remark that here the base function used in \cref{eq: E functions} only depends on the leaf nodes of $\gG$. Thus, we can let $\SpinFunc$ to be the linear combination of different $\gG$ with the same $S_{a_t}$ and $f_{\text{id}}(a_t)$ without introducing additional base function in the decomposition. We use this method to study the spin structure of \ce{Fe4S4(SCH3)4} as the coupling of different iron centers is unknown.

In fact, the number of configurations used to represent a spin-eigen function can be further reduced when considering the $z$-axis spin conservation. As a starting point, we consider how to represent an open-shell singlet with two electrons in the continuous space. The full state of an open-shell singlet can be represented by:
\begin{equation}
    \Phi=\hat{A}\left([\chi^{\uparrow}(\sigma_1)\chi^{\downarrow}(\sigma_2)-\chi^{\uparrow}(\sigma_2)\chi^{\downarrow}(\sigma_1)] \otimes \psi_1(x_1)\psi_2(x_2)\right),
\end{equation}
Then spin part $\chi^{\uparrow}(\sigma_1)\chi^{\downarrow}(\sigma_2)-\chi^{\uparrow}(\sigma_2)\chi^{\downarrow}(\sigma_1)$ is a rank-2 matrix, which requires at least 2 products of one-body functions to represent it. It seems that we need 2 determinants to represent this function in real space. However, one can quickly figure out that the following ans{\"a}tz can represent the open-shell singlet in a standard QMC calculation:
\begin{equation}
    \Phi(x_1,x_2)=\left|\begin{matrix}
        \psi_1(x_1)&\psi_1(x_2)\\
        \psi_2(x_1)&-\psi_2(x_2)\\
    \end{matrix}\right| \label{eq: singlet ansatz}
\end{equation}
This reduction in the number of determinant should be attributed to the  fixed $z$-axis spin quantum number of the walkers in QMC calculations. More concretely, if we explicitly represent the spin part used in \cref{eq: singlet ansatz}, we have:
\begin{equation}
    \Phi(x_1,\sigma_1, x_2, \sigma_2)=\left|\begin{matrix}
        \psi_1(x_1)\left(\chi^{\uparrow}(\sigma_1)+\chi^{\downarrow}(\sigma_1)\right) &\psi_1(x_2)\left(\chi^{\uparrow}(\sigma_2)+\chi^{\downarrow}(\sigma_2)\right)\\
        \psi_2(x_1)\left(\chi^{\uparrow}(\sigma_1)-\chi^{\downarrow}(\sigma_1)\right)&\psi_2(x_2)\left(\chi^{\uparrow}(\sigma_2)-\chi^{\downarrow}(\sigma_2)\right)\\
    \end{matrix}\right|. \label{eq: singlet ansatz full}
\end{equation}
While \cref{eq: singlet ansatz full} represents an open-shell singlet when  $(\sigma_1,\sigma_2)=(\frac{1}{2},-\frac{1}{2})$ or $(-\frac{1}{2},\frac{1}{2})$, it is non-zero when $(\sigma_1,\sigma_2)=(\frac{1}{2},\frac{1}{2})$ or $(-\frac{1}{2},-\frac{1}{2})$, which means that it is not an open-shell singlet in the entire Hilbert space. However, due to the fixed spin coordinate of walkers in conventional QMC calculation, we can still use QMC methods to optimize the ans{\"a}tz in \cref{eq: singlet ansatz full} and derive the correct property of the system.

To extend the above observation, we focus on decomposing the spin part with a specific $z$-axis quantum number, i.e., try to find a decomposition such that:
\begin{equation}
\Theta_S(\sigma_1,\sigma_2,...,\sigma_N)=\sum_tc_t\prod_j\chi_j^t(\sigma_j), \text{ for } \sigma_i \text{ s.t. }\sum_i\sigma_i=\frac{\Nup-\Ndown}{2}. \label{eq: sz conserved decomposition}
\end{equation}
While finding the optimal decomposition for $\SpinFunc$ based on \cref{eq: sz conserved decomposition} is hard to implement, in this note, we provide an easy-to-implement decomposition based on \cref{eq: sz conserved decomposition} and  \cref{eq: ft product}. We first note that there is a conservation law in the $\tFG$:
\begin{equation}
    \tFG=0 \text{ if } \sum_{t}m_{a_t}\neq\frac{\Nup-\Ndown}{2}.
\end{equation}
This is the result of spin-conservation in the CG coefficients. Then, we note that there is another conservation law in $\tEsigma$:
\begin{equation}
    \tEsigma=0\text{ if } \sum_{t}m_{a_t}\neq\sum_i\sigma_i.
\end{equation}
This is because $h^{S,m}_{2S}$ are also the eigenfunction of $\hat{S}_z$. If we ask $\sum_i\sigma_i=(\Nup-\Ndown)/2$, $\tEsigma$ and $\tFG$ would have the same structure of the zero-value element. Then, in \cref{eq: insert identity}, we can just insert $o-1$ identity matrices:
\begin{equation}
\SpinFunc=\sum_{m'_{a_t},m_{a_t}}F^{\gG}_{m'_{a_1},m'_{a_2},...,m'_{a_o}}\left(\prod_{t=1}^{o-1}\delta_{m'_{a_t},m_{a_t}}\right)\tEsigma,
\end{equation}
where $m'_{a_o}$ and  $m_{a_o}$ have to be the same due to the conservation law on $F^{\gG}$ and $E^{\sigma_1\sigma_2...\sigma_N}$. We can follow the same steps as discussed before to derive a decomposition of $\SpinFunc$. The only difference is that the last root of unit $\xi_{S_{a_o}}$ is replaced with $1$, leading to a smaller decomposition of $\Theta_S$.

\section{Spin Functions in Iron-sulfur Clusters}
For simplicity, we use the recursive definition of binary tree to represent the spin function, $\gG=(\gG_L, \gG_R, S_w)$, where $\gG_L$ and $\gG_R$ represents the left subtree and the right subtree of $\gG$, respectively. $S_w$ is the spin quantum number associated with the current node. 
For the \ce{[Fe2S2]^{2+}} and \ce{[Fe2S2(SCH3)4]^{2-}} clusters, we use the CG coefficients corresponding to $\gG=(\frac{5}{2},\frac{5}{2},S)$ as the spin functions. For the \ce{Fe4S4(SCH3)4} cluster, as discussed before, we use a linear combination of the spin functions with different $\gG\in \mathcal{M}_{\text{Fe}}$, where all the $\mathcal{M}_{\text{Fe}}$ is the set of all the binary tree composited by 4 leaf nodes with $S_w=5/2$.
The spin function has the following expression:
\begin{equation}
    \Theta_S(\sigma_1,\sigma_2,...,\sigma_N)=
    \sum_{j_t}\left(\sum_{\gG\in\gM_{\text{Fe}}}c_{\gG}\text{conj}({\ftFG})\right)\ftEsig,
\end{equation}
where $c_{\gG}$ are learnable parameters. As all the binary tree in $\gM_{\text{Fe}}$ have the same leaf nodes, they share the same $\tilde{E}$. Thus we can first sum over $\gM_{\text{Fe}}$ to derive the coefficient used for each configuration, thereby avoiding additional computational burden. 

\section{Comparison with the $S_{+}$ Penalty Method}
\begin{figure}[htb]
    \centering
    \includegraphics[width=\textwidth]{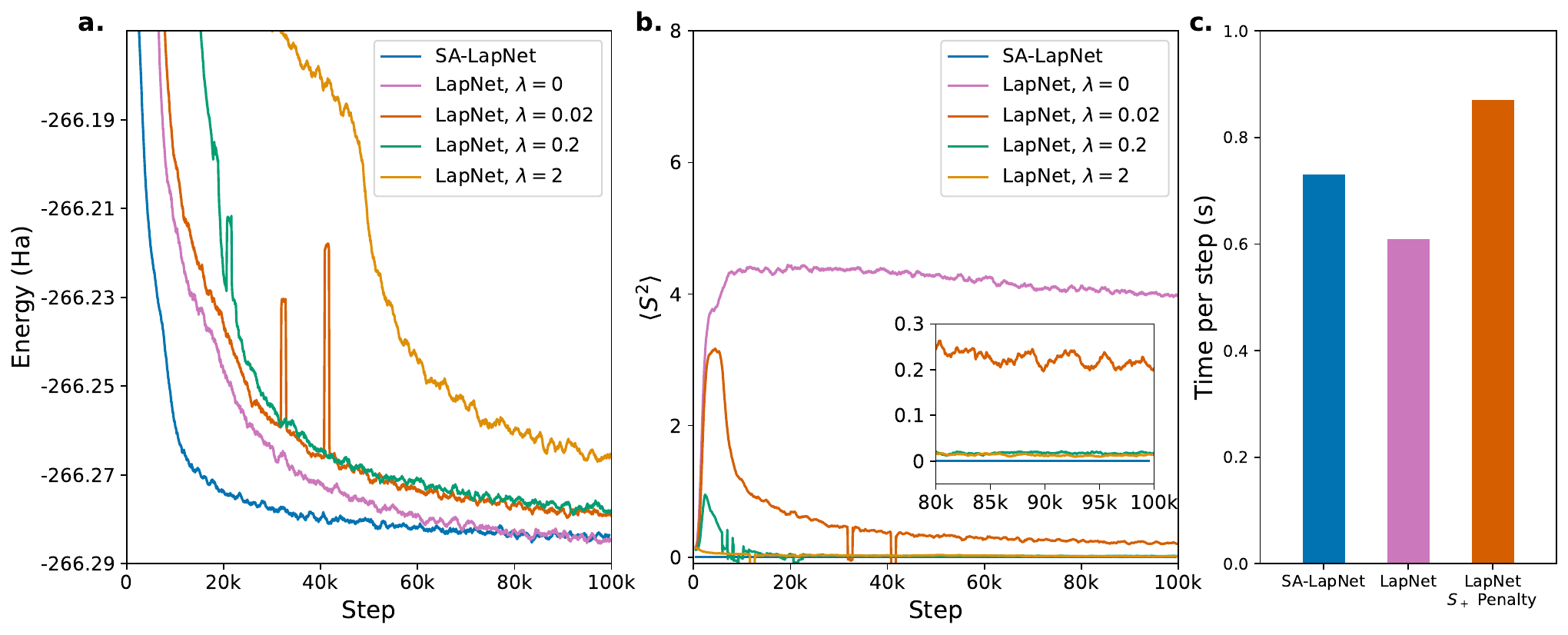}
    \caption{{\bf Comparison between $S_+$ penalty method and SA-LapNet.} We benchmark the performance of SA-LapNet and LapNet with $S_+$ penalty on calculating the $S=0$ state of \ce{[Fe2S2]^2+}. {\bf a}. The energy of different methods during training. $\lambda$ represents the weight used in spin penalty. {\bf b}. The total spin during the training. The inset plot provides a detailed plots of the total spin from 80 000 to 100 000 steps. There is spin-contamination even with a relatively large spin penalty term. {\bf c}. The time per step of different methods.  }
    \label{fig: compare S+}
\end{figure}
In this section, we compare the performance of SA-LapNet and the $S_+$ penalty method on the \ce{[Fe2S2]^{2+}} system. The training curve of different methods are shown in \cref{fig: compare S+}a. Here, $\lambda$ represents the weight of $S_+$ penalty.
Compared with $S_+$ penalty method ($\lambda\neq 0$), SA-LapNet achieves lower absolute energy with a more robust training process. Moreover, as shown by the inset plot of \cref{fig: compare S+}b, there is spin contamination at the end of the training even with a relatively large spin penalty term, demonstrating the importance of enforcing spin symmetry in the wavefunctions. While the LapNet without penalty term ($\lambda=0$) achieves comparable energy result with SA-LapNet, there is spin contamination in this calculation setup as shown in \cref{fig: compare S+}b.  In summary, the SA-LapNet provide the lowest variational energy of $S=0$ state among the 5 calculation step-ups. 
The time per step for each calculation step-ups are plotted in \cref{fig: compare S+}c. The SA-LapNet is slightly slower than the original LapNet implementation due to the additional determinants calculation required by SAAM. However, the SA-LapNet is still faster than the LapNet with a $S_+$ penalty term, demonstrating the efficiency improvement of SAAM.

\renewcommand{\refname}{Supplementary References}
\bibliography{supplement}